\documentclass[onecolumn,preprintnumbers,amsmath,amssymb]{revtex4}

\usepackage{graphicx}
\usepackage{dcolumn}
\usepackage{bm}
\usepackage{epstopdf}
\begin{document}

\title{Electronic and optical properties of the thio-apatites phases Ba$_5$(VS$_{\alpha}$O$_{\beta}$)$_3$X [X=Cl, F, Br, I]: impact of multiple anionic substitution}


\author{Smritijit Sen$^{1}$, Houria Kabbour$^{1, *}$ }

\affiliation{$^1$Univ. Lille, CNRS, Centrale Lille, ENSCL, Univ. Artois, UMR 8181 – UCCS – Unité de Catalyse et Chimie du Solide, Lille, 59655 - Villeneuve d'Ascq Cedex, France.\\
* corresponding author: Houria.KABBOUR@cnrs-imn.fr}

\begin{abstract}
A systematic study of the electronic structure and optical properties of the thio-apatites Ba$_5$(VS$_{\alpha}$O$_{\beta}$)$_3$X (X= Cl, F, Br, I) is carried out through first principles density functional theory simulations. The band gap and properties evolution from fluorine to iodine on fixed O/S ratios, as well as by substituting sulfur (S) for oxygen (O) are discussed. The reduction of the band gap by raising valence band energy levels, with an increasing S/O ratio can also be further modulated by the type of halide in the channels of the structure, thus promoting fine tuning of the band gap region. Defect states also play a crucial role in band gap modulation. Furthermore, the examination of the band edges properties in Ba$_5$(VS$_{\alpha}$O$_{\beta}$)$_3$X compounds suggests they can be potential photocatalysts candidates for the water splitting reaction, with reduced band gaps enabling efficient light-driven reactions, particularly in Ba$_5$(VS$_{\alpha}$O$_{\beta}$)$_3$I. Optical investigations reveal that sulfur doping induces optical anisotropy, enhancing light absorption and offering tailored optical behaviour. These results provide new insights for the design of functional materials in the broad family of apatites.
\vspace{0.5cm}
\end{abstract}
\maketitle
\section{Introduction}
Mixed anionic compounds, characterized by the incorporation of various anionic species within a single structure, constitute a versatile class of materials, renowned for their diverse and possibly enhanced physical and chemical properties \cite{Palazzi1982, Tippireddy2021, Harada2019, Matsumoto2019, Zhang2019, Bacha2024, Kabbour2008, Almoussawi2023l, Sen2021}. Mixed anion compounds are ideal for band gap engineering, which involves controlling and manipulating the band gap between the valence and conduction bands, offering opportunities for optimizing materials for specific device applications, such as solar cells, light-emitting diodes and photodetectors \cite{Kageyama2018, Tripathi2021}. The diverse range of properties found in these compounds, including superconductivity, anion ordering, nonlinear optical behaviour, Ferroelectricity, thermoelectric effects and magnetic characteristics, make them ideal for designing functional materials \cite{Mamouri1994, Sen2023, Li2019, Sen2023jms, Barriga2014, Kamihara2008, Sen2020, Yashima2009, Valldor2014}. Theoretical studies at the microscopic level and prior to synthesis will enhance our understanding of these compounds. Theoretical calculations indicate that the valence band maximum is primarily influenced by hybridized anion orbitals. Specifically, less electronegative anions raise the valence band maximum, providing substantial opportunities for band gap engineering through anionic manipulation. Beyond the well-known photocatalytic activity of oxynitrides for water-splitting in the visible range, recent studies have shown that some oxysulfides, such as Y$_2$Ti$_2$O$_5$S$_2$ and LaOInS$_2$, exhibit promising stability and activity \cite{Miyoshi2021, Wang2019, Kabbour2004, Kabbour2020}. Additionally, low-dimensional thiovanadate compounds have emerged as a fertile ground for band gap engineering, exploiting various S/O ratios \cite{Almoussawi2020, Almoussawi2023, Nicoud2019}. The heteroleptic environment, characterized by at least two types of anions around the cation, enhances local acentricity and polarity in polar structures, thereby improving photoelectric response by facilitating better charge carrier separation \cite{Bacha2022}. Apatite serves as another prime example of mixed anionic compounds, renowned for its wide range of applications. Apatites M$_{10}$(AO$_4$)$_6$X$_2$ (where M and A represent cationic sites, A is in unconnected tetrahedral AO$_4$, and X is a halide in the channel) are common compounds with diverse practical applications, including fertilizers, food additives, bioceramics, catalysts, and fluorescent host materials. Notably, Pb$_5$(VO$_4$)X (X = F, Cl, Br, I) apatites have recently been identified for their light-absorption properties, opening new perspectives among Apatite materials. The iodide phase, with a band gap in the visible range (2.7 eV), owes this property to the contribution of I 5p states at the top of the valence band, making it suitable as a photo-anode for photoelectric conversion \cite{Nakamura2020}. Exploratory synthesis has led to the discovery of the oxysulfide Ba$_{10}$S(VO$_3$S)$_6$, which exhibits a supercell similar to apatite but lacks halide anions, featuring sulfur in both the thiovanadates and the channels, resulting in a band gap of 2.25 eV \cite{Almoussawi2020}. To develop materials active under visible light by inserting sulfide anions into oxy-halide apatites, a new family of halide-thioapatites, Ba$_5$(VS$_{\alpha}$O$_{\beta}$)$_3$X (X = Cl, F, Br, I), containing thiovanadate groups instead of vanadate groups has been discovered \cite{Almoussawi2023l}. Ba$_5$(VS$_{\alpha}$O$_{\beta}$)$_3$X (X = Cl, F, Br, I) compounds represent a unique class of materials with intriguing optical properties owing to their rich anionic lattice (halide X$^-$, O$^{2-}$ and $^{2-}$). composition. These compounds offer the potential for tuning their electronic and optical behaviour by varying the halogen component (X) and the ratio of sulfur to oxygen. Such tunability makes them promising candidates for applications in optoelectronics, photocatalysis, and luminescent devices. The synthesis, structural characteristics, and optical properties of these compounds to explore their potential were reported in \cite{Almoussawi2023l}. Experimentally determined structures are however available for X= F and Cl, and not for the other halides which could not be investigated or could not be fully solved due to insufficient single crystal quality (in the case of Iodide).\\

In this work, we explore the electronic and optical properties of Ba$_5$(VS$_{\alpha}$O$_{\beta}$)$_3$X (X = Cl, F, Br, I) through our first principles density functional theory simulations. We also uncover the potential of Ba$_5$(VS$_{\alpha}$O$_{\beta}$)$_3$X systems as efficient photocatalysts for water splitting, facilitated by their reduced band gaps and visible light absorption through sulfur doping.

\begin{figure}[h!]
\centering
\includegraphics [height=5cm,width=12.5cm]{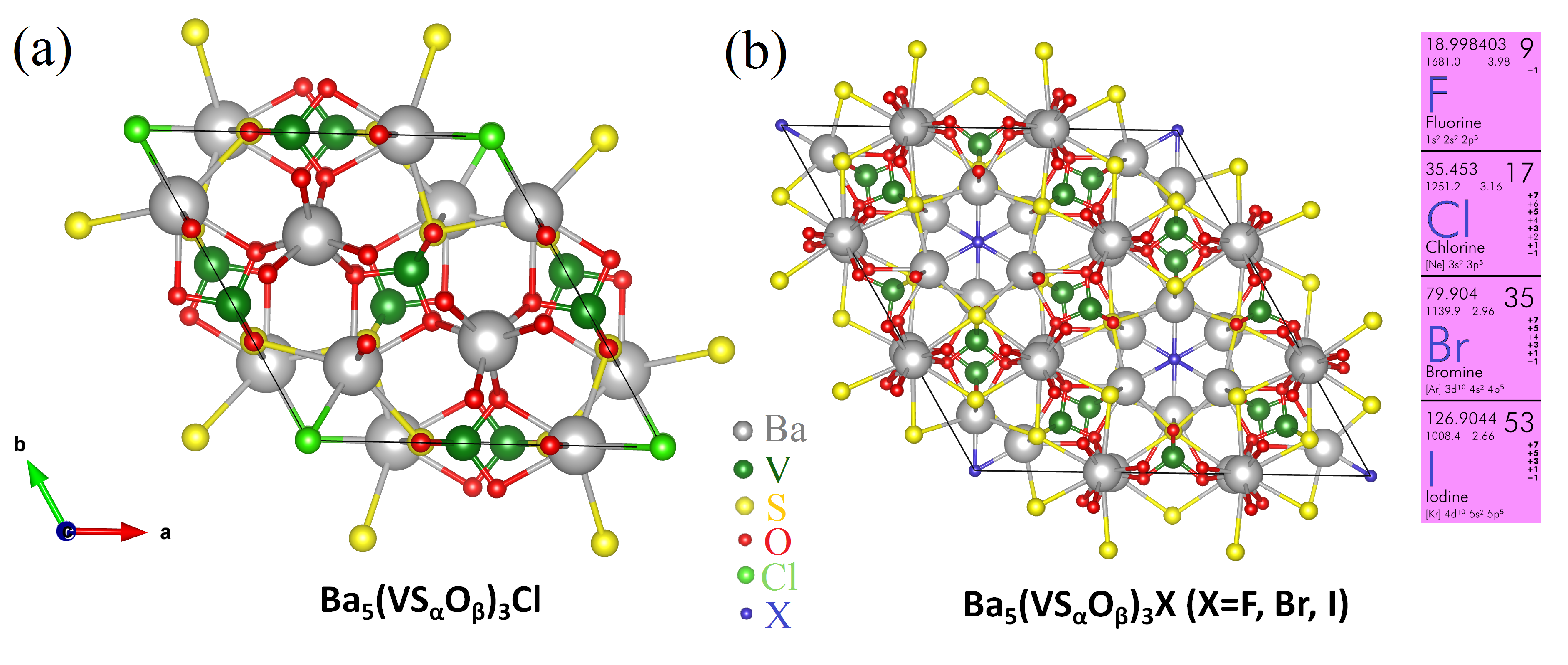}
\caption{Crystal structure of (a) Ba$_5$(VS$_{\alpha}$O$_{\beta}$)$_3$Cl (b) Ba$_5$(VS$_{\alpha}$O$_{\beta}$)$_3$X (X= F, Br, I).}
\label{str}
\end{figure}
\section{Crystal structure and computational method}
Experimental lattice parameters and atomic positions of Ba$_5$(VS$_{\alpha}$O$_{\beta}$)$_3$X (X= Cl, F) are used as the input of our first principles density functional theory calculations \cite{Kabbour2020}. We perform full geometry optimization (relaxation of lattice parameters as well as atomic positions) for the Ba$_5$(VS$_{\alpha}$O$_{\beta}$)$_3$X (X= F, Br, I) crystal structure. It is to be mentioned here that experimental crystallographic data is only available for Ba$_5$(VS$_{\alpha}$O$_{\beta}$)$_3$X (X= Cl, F). The crystal structures were visualized by
the VESTA software \cite{Momma2008} and presented in FIG.\ref{str}. Our first principles density functional theory calculations are performed using VASP (Vienna Ab initio Simulation Package) which employs the plane wave basis set to describe the electronic wave-functions \cite{Kresse1993, Bloch1994, Kresse1996}. For structure optimization, electronic structure evaluation as well as for optical property calculations, the exchange-correlation functional has been treated under the Meta-Generalized Gradient Approximation (meta-GGA) SCAN (strongly constrained and appropriately normed) \cite{Sun2015} functional. The wave functions were expanded in a plane wave basis with an energy cut-off of 550 eV and the energy tolerance of the self consistent calculations is set to 10$^{-6}$ eV for SCAN functional. In the structural optimizations, the atomic positions are relaxed until the Hellmann-Feynman forces are less than 0.001 eV/\AA. The sampling of the Brillouin zone is done using a $\Gamma$ centered $3\times3\times3$  Monkhorst-Pack grid. The linear optical properties can be procured from the frequency-dependent dielectric function $\epsilon (\omega)=\epsilon_1(\omega)+i\epsilon_2(\omega)$, where $\epsilon_1(\omega)$, $\epsilon_2(\omega)$ are the real and imaginary parts of the dielectric function and $\omega$ is the incident photon frequency. The imaginary part $\epsilon_2(\omega)$ of the dielectric function can be evaluated from the momentum matrix elements between the occupied and unoccupied wave functions \cite{Gajdos2006}. The real and imaginary parts of the analytical dielectric function are connected by the Kramers-Kronig relationship \cite{Dresselhaus1999}. Frequency dependent absorption, electron energy loss spectra (EELS), reflectivity, refractive index, etc., can be obtained from the real and imaginary parts of the frequency dependent dielectric function \cite{Saha2000, Luo2015, Peng2016}.
\section{Results and discussion}
\subsection{Electronic properties}
\begin{figure}[h!]
\centering
\includegraphics [height=10.5cm,width=7cm]{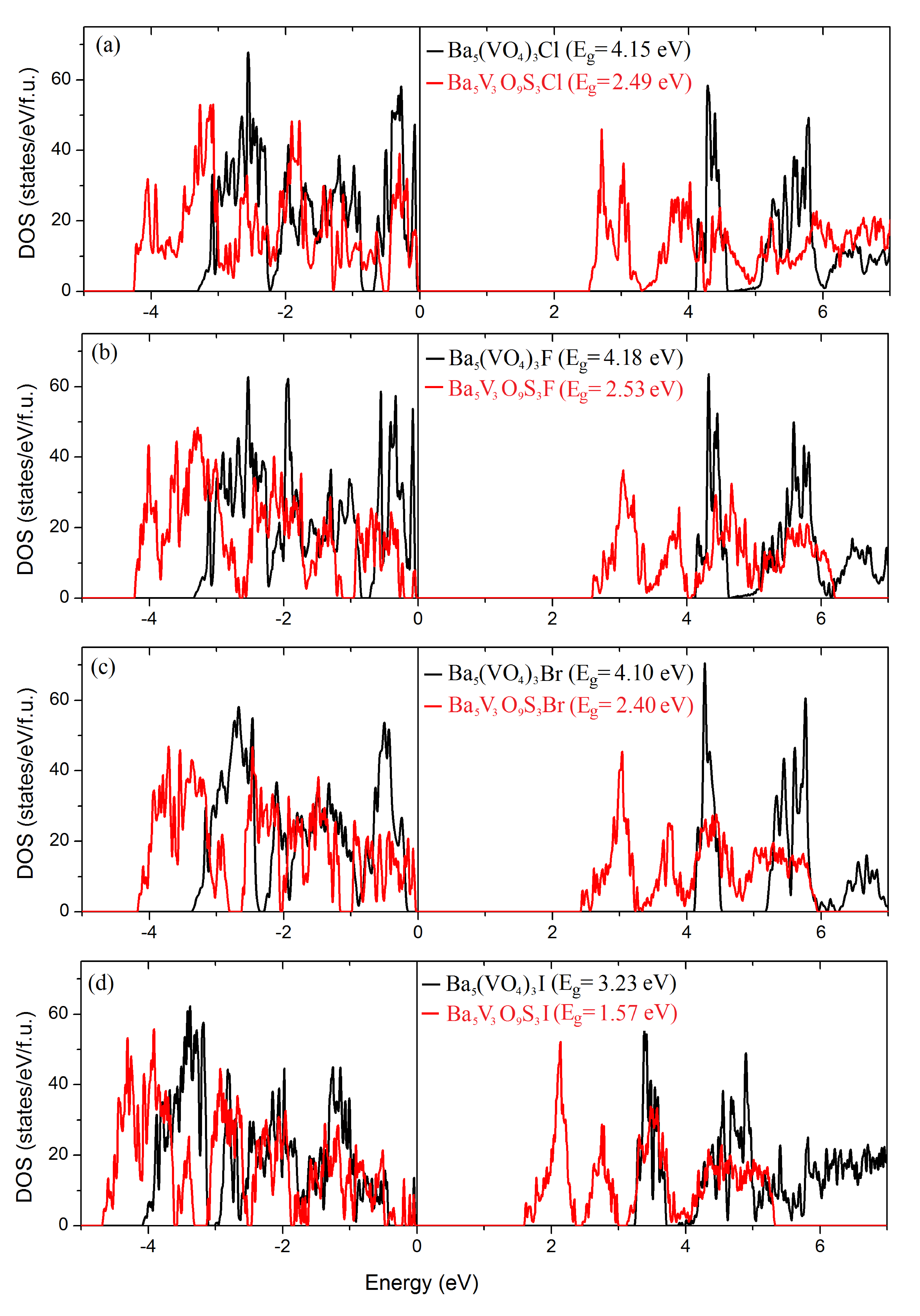}
\caption{Calculated density of states of Ba$_5$(VS$_{\alpha}$O$_{\beta}$)$_3$X (X= Cl, F, Br, I).}
\label{dos}
\end{figure}
\begin{figure}[h!]
\centering
\includegraphics [height=5cm,width=6cm]{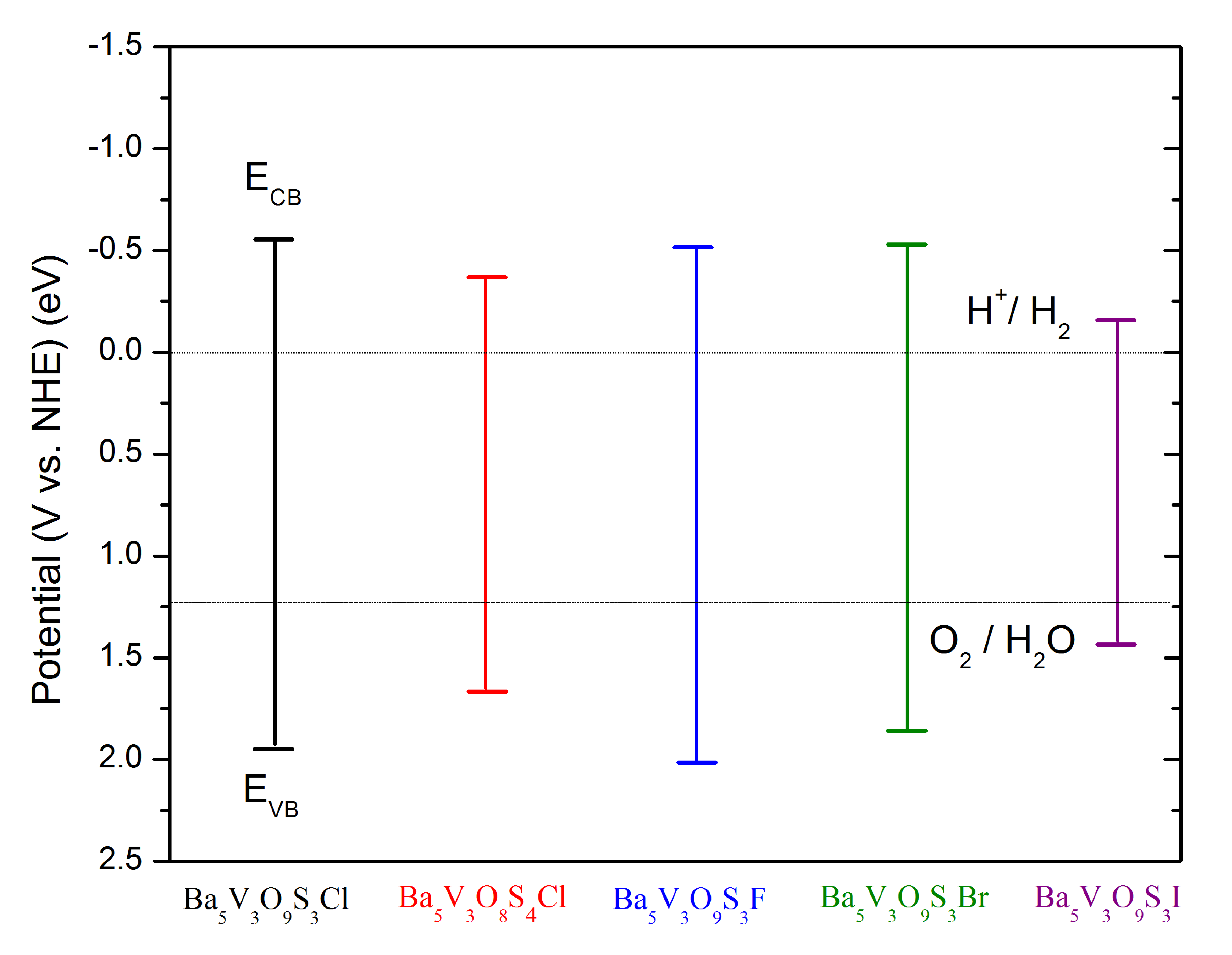}
\caption{Calculated band edge potentials of Ba$_5$(VS$_{\alpha}$O$_{\beta}$)$_3$X (X= Cl, F, Br, I).}
\label{pcat}
\end{figure}
\begin{table}[h!]
  \begin{center}
\caption{Calculated band gap of Ba$_5$(VS$_{\alpha}$O$_{\beta}$)$_3$X (X= Cl, F, Br, I). Experimentally measured values of band gaps are in the brackets.}
    \vspace{2mm}
    \label{bandgap}
    \begin{tabular}{c|c|c|c|c|c|c|c} 
     \hline
     System & Band-gap (eV) & System & Band-gap (eV) & System & Band-gap (eV) & System & Band-gap (eV)\\
    \hline
     Ba$_5$(VO$_4$)$_3$Cl & 4.15 & Ba$_5$(VO$_4$)$_3$F & 4.18 & Ba$_5$(VO$_4$)$_3$Br & 4.10 & Ba$_5$(VO$_4$)$_3$I & 3.23 \\
     \hline
     Ba$_5$V$_3$O$_9$S$_3$Cl & 2.49 (2.39*)& Ba$_5$V$_3$O$_9$S$_3$F & 2.53 (2.48*)& Ba$_5$V$_3$O$_9$S$_3$Br & 2.40 & Ba$_5$V$_3$O$_9$S$_3$I & 1.57  \\
     \hline
     Ba$_5$V$_3$O$_8$S$_4$Cl & 2.03 (2.11*) &-  &-  & - & - & - & - \\ 
     \hline
      \end{tabular}
  \end{center}
   *Reference \cite{Almoussawi2023}
\end{table}

In this section, we will discuss the electronic structure of Ba$_5$(VS$_{\alpha}$O$_{\beta}$)$_3$X (X= Cl, F, Br, I) highlighting the band gap and photo-catalytic properties. In FIG.\ref{dos}, we depict our calculated density of states for Ba$_5$(VS$_{\alpha}$O$_{\beta}$)$_3$X (X= Cl, F, Br, I). It is noteworthy to remind the electronegativities evolution among the studied halide anions with $\chi$(F)= 4.0$>\chi$(Cl)= 3.0$>\chi$(Br)= 2.8$>\chi$(I)= 2.5 in Pauling scale, despite which the Cl has a greater electronic affinity than F due to a reduced electronic repulsion in Cl related to its greater size. It is very clear from our calculation that as we move from F to I, the band gap of Ba$_5$(VO$_4$)$_3$X (X= F, Cl, Br, I) decreases as expected. When a highly electronegative element like fluorine is part of the compound, it tends to pull electron density towards itself more strongly. This strong attraction increases the energy difference between the valence band and the conduction band, leading to a larger band gap. As we move to less electronegative elements like iodine, this effect diminishes, resulting in a smaller band gap. Moreover, the atomic size and polarizability of the halogens increase from fluorine to iodine. Larger atoms like iodine have more diffuse electron clouds, which can lead to more overlap between orbitals and reduce the band gap. The increased polarizability of iodine also means that the electronic environment is more easily distorted, which can facilitate electronic transitions and thus lower the band gap. Therefore, by replacing halogens, we can tune the band gap in Ba$_5$(VO$_4$)$_3$X (X= F, Cl, Br, I). Next we will discuss the effect of S doping in O site. We can clearly see that the band gap decreases as we replace one O by S in Ba$_5$(VO$_4$)$_3$X (X= F, Cl, Br, I). S has a lower electronegativity and higher atomic size compared to O. When S atoms replace O atoms in the lattice, the energy levels of the valence band (mainly constituted by p orbitals of S and O) are raised. This is because S-3p orbitals are higher in energy compared to O-2p orbitals, resulting in a smaller energy difference between the valence band and conduction band. In other words, S-3p orbitals overlap differently with the metal cations compared to O-2p orbitals. This altered overlap can reduce the band gap by affecting the band structure and electronic transitions within the material. We have presented our calculated band structures of Ba$_5$(VS$_{\alpha}$O$_{\beta}$)$_3$X (X= Cl, I) in the supplementary materials. We also observed that by increasing the S/O ratio, one can further reduce the band gap in Ba$_5$(VO$_4$)$_3$Cl (see supplementary material for the calculated density of states for Ba$_5$V$_3$O$_8$S$_4$Cl). However, another key factor behind the modulation of the band gap in Ba$_5$(VS$_{\alpha}$O$_{\beta}$)$_3$X (X= Cl, F, Br, I) is the presence of defects states, as quite evident from the experimental crystallographic data \cite{Almoussawi2023}. In TABLE\ref{bandgap}, we present the calculated band gaps of Ba$_5$(VS$_{\alpha}$O$_{\beta}$)$_3$X (X= Cl, F, Br, I) and compare them with the available experimentally measured band gaps. All the Ba$_5$(VS$_{\alpha}$O$_{\beta}$)$_3$X (X= Cl, F, Br, I) systems show an indirect bang gap. Our calculated band gaps are quite consistent with the experimental results. As mentioned earlier, Ba$_5$(VS$_{\alpha}$O$_{\beta}$)$_3$X (X= Br, I) are not yet synthesized. Our calculations suggest that one can obtain band gap as low as 1.57 eV in Ba$_5$V$_3$O$_9$S$_3$I. However, the role of defect states in the modulation of band gap can only be explored in Ba$_5$(VS$_{\alpha}$O$_{\beta}$)$_3$X (X= Br, I) once we have the experimental crystallographic data. Now we discuss the potential of Ba$_5$(VS$_{\alpha}$O$_{\beta}$)$_3$X (X= Cl, F, Br, I) as photocatalysts. Photocatalytic water splitting in Ba$_5$(VS$_{\alpha}$O$_{\beta}$)$_3$X (X= Cl, F) is already been explored experimentally \cite{Almoussawi2023}. Photocatalytic water splitting involves using light energy to separate water molecules into hydrogen and oxygen. The fundamental principle behind this process is the use of a photocatalyst that absorbs photons from a light source, typically sunlight, and uses that energy to drive the water-splitting reaction. When the photocatalyst absorbs light energy, it generates electron-hole pairs. These excited electrons and holes then engage in redox reactions, which facilitate the splitting of water. For an efficient photocatalyst, the band gap should be in the visible light region ($1.6-3.2$ eV). It is evident that S doping in Ba$_5$(VO$_4$)$_3$X (X= F, Cl, Br, I), reduces the band gap, bringing it into the visible light region. The water-spitting process and redox ability of Ba$_5$(VS$_{\alpha}$O$_{\beta}$)$_3$X (X= Cl, F, Br, I) are determined by the valence and conduction band edge potentials, which can be estimated from the following mathematical expressions:
\begin{equation}
    E_{VB}=\chi-E_e+E_g/2
\end{equation}
\begin{equation}
    E_{CB}=E_{VB}-E_g
\end{equation} 
where $E_{VB}$ and $E_{CB}$ are the valence and conduction band edge potentials, respectively. $\chi$ is the Mulliken electronegativity of Ba$_5$(VS$_{\alpha}$O$_{\beta}$)$_3$X (X= Cl, F, Br, I). In FIG.\ref{pcat}, we depict the valence and conduction band edge potentials of Ba$_5$(VS$_{\alpha}$O$_{\beta}$)$_3$X (X= Cl, F, Br, I) along with the water splitting and oxidation potentials for reference. For an efficient water splitting photocatalytic material, $E_{VB}>1.23V$ and $E_{CB}<-0.33V$. From FIG.\ref{pcat}, it is clear that Ba$_5$V$_3$O$_9$S$_3$I is the most efficient water splitting photocatalytic material as compared to the other four.
\subsection{Optical properties}
\begin{figure}[h!]
\centering
\includegraphics [height=8.5cm,width=11cm]{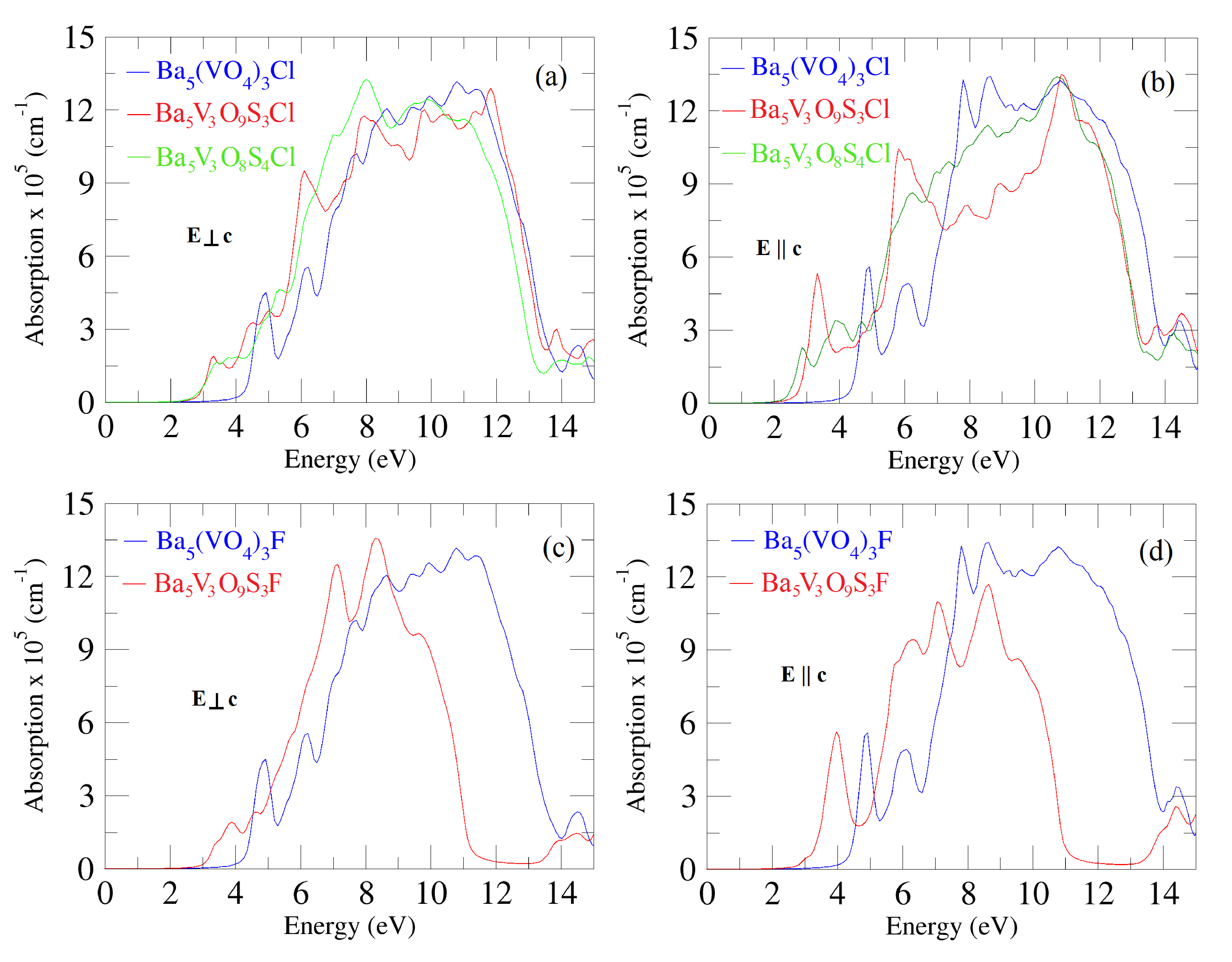}
\caption{Calculated absorption of Ba$_5$(VS$_{\alpha}$O$_{\beta}$)$_3$X (X= Cl, F).}
\label{abs1}
\end{figure}
\begin{figure}[h!]
\centering
\includegraphics [height=8.5cm,width=11cm]{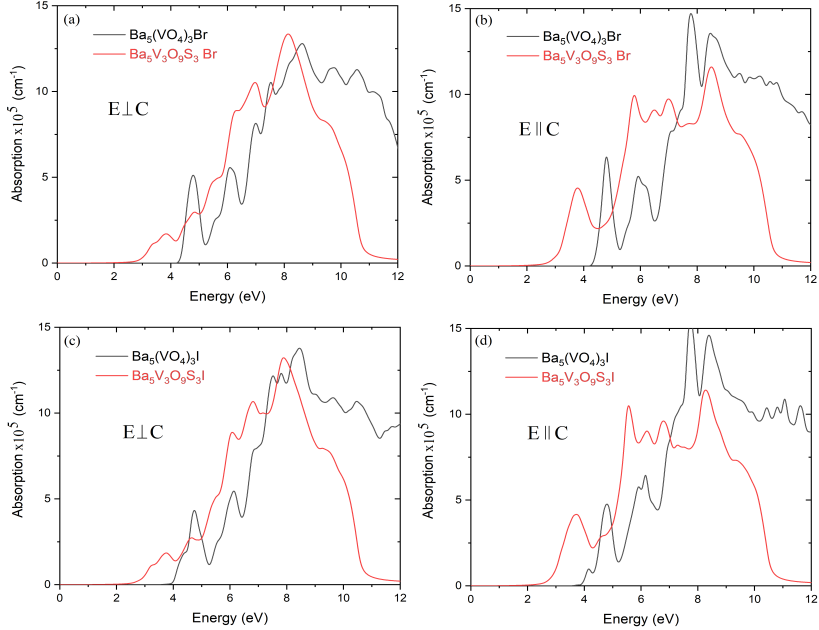}
\caption{Calculated frequency dependent absorption of Ba$_5$(VS$_{\alpha}$O$_{\beta}$)$_3$X (X= Br, I).}
\label{abs2}
\end{figure}
\begin{figure}[h!]
\centering
\includegraphics [height=6.5cm,width=14.5cm]{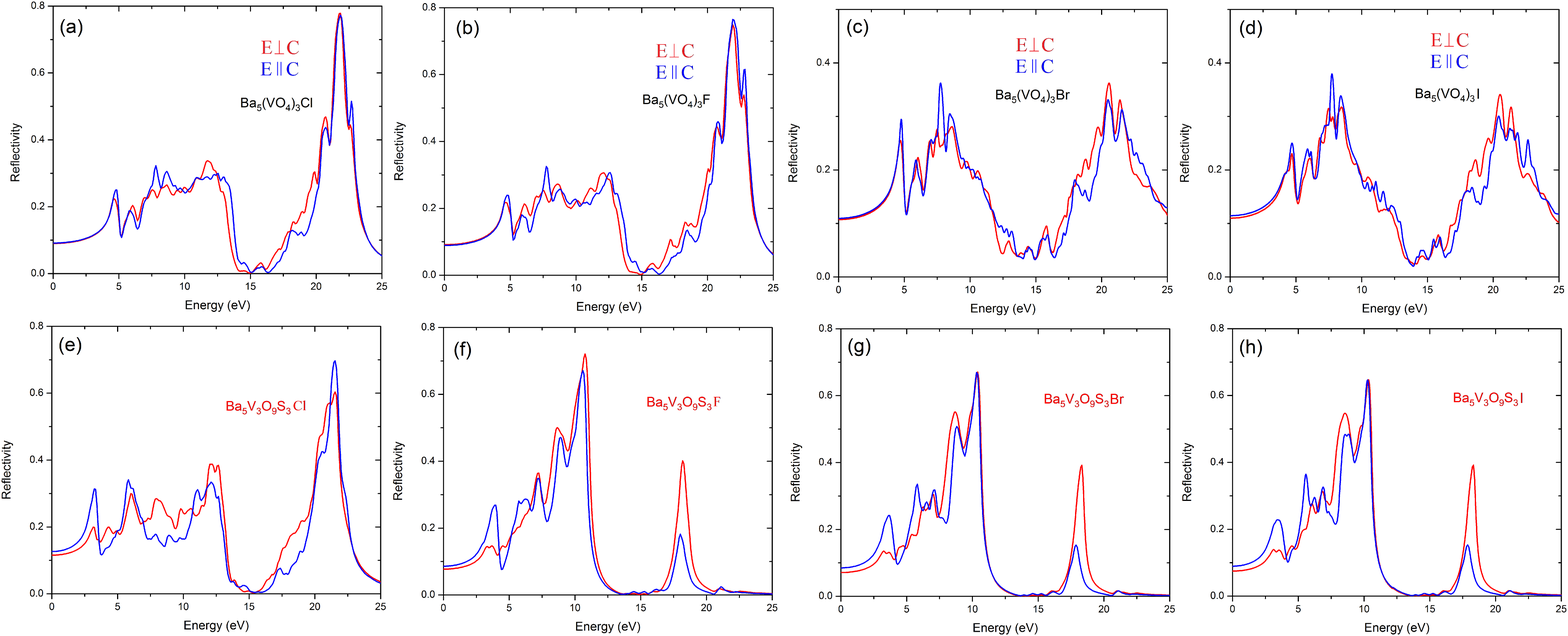}
\caption{Calculated frequency dependent reflectivity of Ba$_5$(VS$_{\alpha}$O$_{\beta}$)$_3$X (X= Cl, F, Br, I).}
\label{ref}
\end{figure}
\begin{figure}[h!]
\centering
\includegraphics [height=6.5cm,width=14.5cm]{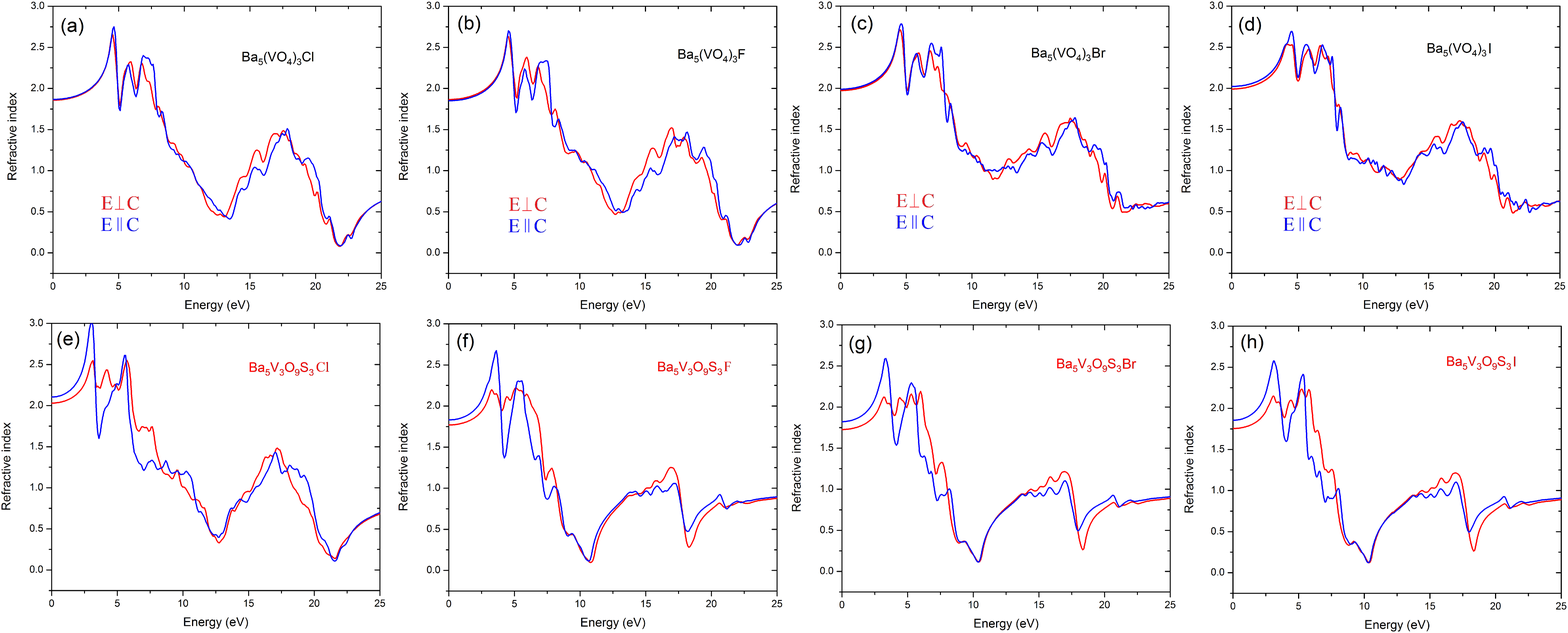}
\caption{Calculated frequency dependent refractive index of Ba$_5$(VS$_{\alpha}$O$_{\beta}$)$_3$X (X= Cl, F, Br, I).}
\label{ri}
\end{figure}
 The optical properties of these materials are examined through light-matter interactions, where incident photon energy absorbed by valence electrons initiates intra- or inter-band transitions \cite{Vu2020, Fox2002}. However, an ideal optoelectronic material must facilitate inter-band transitions, as these are crucial for maximum incident energy absorption, which is determined by the electronic band gap. The dielectric function, intrinsic to the electronic band structure of a semiconductor, dictates its response to incident light (electromagnetic radiation) through the interaction of electrons and photons \cite{Hoat2019, Yuan2014}. Various optical parameters are computed for Ba$_5$(VS$_{\alpha}$O$_{\beta}$)$_3$X (X= Cl, F, Br, I), including absorption spectra, energy loss spectra, reflectivity and refractive index. In FIG.\ref{abs1} and FIG.\ref{abs2}, we depict our calculated absorption for Ba$_5$(VS$_{\alpha}$O$_{\beta}$)$_3$X (X= Cl, F) and Ba$_5$(VS$_{\alpha}$O$_{\beta}$)$_3$X (X= Br, I) respectively. For all the systems, we present absorption spectra for parallel as well as perpendicular light polarizations. The absorption spectra clearly indicate that replacing O with S in Ba$_5$(VO$_4$)$_3$X (X= F, Cl, Br, I) significantly enhances its absorption in the visible light range. This modification suggests that the S doping effectively tunes the material's optical properties, making it more responsive to visible light. Another noteworthy observation is that doping with S substantially increases the optical anisotropy of the material. This enhancement suggests that S doping not only affects the absorption characteristics but also influences the directional dependence of the material's optical properties, potentially leading to improved performance in applications requiring controlled light propagation. The absorption spectra reveal a distinct peak when the incident photon is aligned parallel to the optical c-axis for Ba$_5$(VS$_{\alpha}$O$_{\beta}$)$_3$X (X= Cl, F, Br, I) in the visible light range. However, the position of this peak varies at different energy levels depending on the specific halogen present in Ba$_5$(VS$_{\alpha}$O$_{\beta}$)$_3$X (X= Cl, F, Br, I). This variation indicates that each halogen (F, Cl, Br, I) influences the material's optical response differently, leading to shifts in the peak position and suggesting a tailored interaction between the system and the incident light. Further S doping causes the absorption peak of Ba$_5$V$_3$O$_8$S$_4$Cl to decrease in intensity and shift towards longer wavelengths (red shift), indicating that the S/O ratio significantly influences the optical properties. Additionally, the presence of defect states plays a crucial role in controlling the optical properties of Ba$_5$(VS$_{\alpha}$O$_{\beta}$)$_3$X (X= Cl, F, Br, I). In FIG.\ref{ref} and FIG.\ref{ri}, we depict our calculated frequency dependent reflectivity and refractive index of Ba$_5$(VS$_{\alpha}$O$_{\beta}$)$_3$X (X= Cl, F, Br, I), respectively. It is quite evident form FIG.\ref{ref} and FIG.\ref{ri} that S doping introduces optical anisotropy in the frequency-dependent reflectivity and refractive index in the visible light region for these materials. The introduction of S doping leads to a general decrease in static reflectivity and refractive index, except in the case of Ba$_5$V$_3$O$_9$S$_3$Cl and Ba$_5$V$_3$O$_8$S$_4$Cl (see supplementary material). A decrease in static reflectivity often indicates an increase in light absorption, as less light is being reflected from the material's surface. This suggests that the doped material is more effective at absorbing incident light, which could be advantageous for applications like photovoltaics or photodetectors.
\section{Conclusion}
In this article, we have investigated more into depth the electronic structure and optical properties of the thio-Apatites family of compounds Ba$_5$(VS$_{\alpha}$O$_{\beta}$)$_3$X (X= Cl, F, Br, I). The later is derived from the broad interest Apatite-type by replacement of the oxide polyanionic groups, vanadates VO$_4$ here, by mixed anion polyanionic groups, the thiovanadates V(O$_{4-x}$S$_x$). It allowed a better comprehension of the impact of anionic substitution on the electronic structure (and optical properties) with two types of anionic sites substituted and a triple anionic lattice :  O$^{2-}$ and $^{2-}$ forming the thiovanadates and the halide (X$^-$) within the channels. The substitution within the channels contributes to reduce the band gap from Fluorine to Iodine following the variation of electronegativities and atomic sizes of the halides. Meanwhile, the substitution of O by S contributes also to reduce the band gap by raising the energy levels of the valence band. The two concomitant effects (at the halide site and at the chalcogenide site) allow a fine tuning of the band gap. We also found that the defect states played a significant role in modulating the band gap of these materials. On another hand, the determined valence and conduction band edges potentials suggest that Ba$_5$(VS$_{\alpha}$O$_{\beta}$)$_3$X (X= Cl, F, Br, I) may act as photocatalysts for the water splitting reaction. The reduced band gaps, particularly in Ba$_5$(VS$_{\alpha}$O$_{\beta}$)$_3$I, bring them into the visible light region, making them suitable candidates for light-driven chemical reactions. Furthermore, we observe that sulfur doping introduces optical anisotropy in the frequency-dependent reflectivity and refractive index, particularly in the visible light region, as well as a decrease in static reflectivity. These alterations indicate a potential for enhanced light absorption and tailored optical behaviour in sulfur-doped materials, which could have implications for various optoelectronic applications. Overall, we show the versatile and tuneable properties of Apatites Ba$_5$(VS$_{\alpha}$O$_{\beta}$)$_3$X (X= Cl, F, Br, I) based on an original triple anion lattice, which open new perspectives to design functional materials.

\section{Acknowledgments}
This work was supported by the I-Site (ANR-16-IDEX-0004 ULNE) and by the ANR (project ANR-21-CE08-0051). The regional computational cluster (Mesocentre-Lille) is thanked for providing computational resources.

\section{Declaration of competing interest}
The authors declare that they have no known competing financial interests or personal relationships that could have appeared to influence the work reported in this paper.

\section{Data availability}
Data will be available upon considerable request to the corresponding author.

\end{document}